# Temperature Dependence of Magnetocrystalline Anisotropy in Itinerant Ferromagnets

Daisuke Miura* and Akimasa Sakuma

*Department of Applied Physics, Tohoku University, Sendai 980-8579, Japan*

We theoretically investigated magnetocrystalline anisotropy (MA) at a finite temperature $T$ in ferromagnetic metals. Assuming a Rashba-type ferromagnet with uniaxial MA, we defined the MA constants $K_\mathrm{u}(T)$ derived from several different concepts. Our purpose was to examine the equality between them and to confirm a power law between $K_\mathrm{u}(T)$ and magnetization $M(T)$ in the form of $K_\mathrm{u}(T)/K_\mathrm{u}(0) = [M(T)/M(0)]^\alpha$. We demonstrate that $\alpha$ equals 2 in the itinerant-electron limit and increases with the localized feature of electrons passing through $\alpha = 3$, predicted for the single-ion MA in spin models.

Magnetocrystalline anisotropy (MA) is an important factor affecting the performance of magnetic materials; thus, understanding and controlling MA are critical requirements in the field of magnetics [1]. Recently, MA, especially its temperature dependence, has been actively investigated for rare-earth (RE) permanent magnets [2], and the most recent studies have also focused on RE-free permanent magnets [3]. In general, the temperature dependence of the MA constants (MACs) of RE magnets is complex [4]. However, the basic mechanism of this dependence has been clearly understood on the basis of a local moment picture, where the MACs of a material are defined in terms of the magnetization-direction dependence of the Helmholtz energy density of the material, which describes the degree of MA [5–7]. In this picture, a semi phenomenological interpretation can be found by focusing on the interaction between the crystalline-electric field and the non-spherical localized-electron cloud at each site. For example, we have demonstrated that the temperature dependence of the MACs of Nd-Fe-B magnets can be accurately reproduced by a power law of the Nd-site moment [8,9]. Here, a power law [8] extended from the conventional Akulov–Zener–Callen–Callen (AZCC) law [10–12] must be used. In addition to such semi-phenomenological treatment, theoretical investigations for MA in finite temperatures within the local moment picture have been performed on the basis of the ligand-field theory [7,13–17], the Landau–Lifshitz–Gilbert equation [18–20], the Monte-Carlo method [21–24], and other techniques [7,25,26]. However, compared with the large number of theoretical studies on *localized* electron systems, there are few reports on *itinerant* systems, such as RE-free permanent magnets. In fact, a trial of the theoretical reproduction of the Curie temperature of iron is still ongoing [27]. Moreover, for the MA of Nd-Fe-B magnets, the contribution from the Fe sublattice is not understood completely [8,28,29]. In this regard, Y-Fe-B magnets are often considered because Y is non-magnetic, and several theoretical examinations of its MA have been reported [28,29].

Furthermore, for *itinerant* electron systems such as ferromagnetic metals, researchers have attempted to understand the temperature-dependent MA on the basis of the temperature dependence of the magnetization. Thiele et al. [30] and Okamoto et al. [31] experimentally showed that in FeNiPt alloys and FePt, respectively, the first-order MAC $K_\mathrm{u}(T)$ is roughly proportional to $M(T)^2$, where $M(T)$ is the saturation magnetization at a temperature $T$. This second power law does not obey the prediction of $M(T)^3$ as per the AZCC theory, which suggests that the local moment picture is not appropriate [32]. In such a situation, the coherent potential approximation (CPA) theory in the disordered local moment (DLM) picture based on the spin density functional theory has provided a quantitative description of the finite temperature magnetism in itinerant ferromagnets [33–36]. The DLM-CPA theory has succeeded at reproducing the second power law in FePt [37,38], and it has also shown that other ferromagnetic metals obey the second power law for FePd [38,39], CoPt, MnAl, and FeCo [38], from the first principles. Recently, the DLM-CPA theory has been also employed to evaluate the temperature dependence of important factors in spintronic devices, such as the half metallicity of Co$_2$MnSi [40] and the Gilbert damping constant of Fe and FePt [41].

Very recently, we have demonstrated, on the basis of the DLM-CPA theory, that the zero-field transverse spin susceptibility $\chi_\perp$ of Rashba-type ferromagnets is almost constant in the range of $T < T_\mathrm{C}$ (the Curie temperature), and that $\chi_\perp$ obeys the Curie–Weiss law for $T > T_\mathrm{C}$ [42]. As suggested therein, this result indicates $K_\mathrm{u}(T) \propto M(T)^2$ rather than $M(T)^3$. This and the aforementioned first-principles results imply that the second power law is widely valid in ferromagnetic metals with a uniaxial MA. Although we have referred to a relation between magnetic susceptibility and an MAC, we have not yet discussed this in detail. In this letter, using the Rashba-type ferromagnetic model [42–46], we examine this relation and demonstrate its validity within the DLM-CPA theory. In addition, we consider the power law in ferromagnetic metals.

We now consider only uniaxial-MA ferromagnetic systems. Let the z and x axes in the Cartesian coordinates be the easy and hard axes, respectively. We introduce three targets examined as a uniaxial MAC: (1) $K_\mathrm{u}^\mathrm{MP}$ evaluated from the magnetization process, (2) $K_\mathrm{u}^\mathrm{EXF}$ defined by the angle dependence of the Helmholtz energy density on the rotation of exchange fields (EXFs), and (3) $\widetilde{K}_\mathrm{u}^\mathrm{EXF}$ obtained by a perturbative expansion of $K_\mathrm{u}^\mathrm{EXF}$ with respect to a spin-orbit interaction (SOI).

*Definition of* $K_\mathrm{u}^\mathrm{MP}$. The relationship between $\chi_\perp(T)$ and an MAC can be determined by applying the definition of $\chi_\perp(T)$ to the Sucksmith–Thompson equation as follows [42,47,48].

$$K_\mathrm{u}^\mathrm{MP}(T) := \frac{\mu_0 M(T)^2}{2\chi_\perp(T)}, \tag{1}$$

where $\mu_0$ is the permeability in vacuum. We assume that the higher-order MACs are negligible; thus, Eq. (1) cannot be used for complex-MA materials such as RE permanent magnets [4]. This expression clarifies the temperature dependences of $M(T)$ and the MAC, that is, we can interpret that $\chi_\perp(T)$ plays a role in the correction for the second power law. Although Eq. (1) is a phenomenological expression, for a given Hamiltonian, one can microscopically obtain $\chi_\perp(T)$ within the DLM-CPA theory by numerically solving the integral equation described in our previous work [42]. In this study, we employ the same model. Accordingly, we consider the electron system described by the tight-binding Rashba Hamiltonian [49] with the EXF on a two-dimensional square lattice having a lattice constant $a$ [42,44,46]. This Hamiltonian is defined by

$$\mathcal{H}\{e_j\} := \mathcal{H}_0\{e_j\} + \mathcal{V}. \tag{2}$$

Here, local EXFs are defined at each site in a crystal. The directions of the EXFs are represented by a configuration $\{e_j\} := (e_1, e_2, \ldots, e_N)$, where $e_j$ is the unit vector representing the direction of the EXF located at a lattice vector $R_j$, and $N$ is the total number of the sites. The first and second terms in Eq. (2) are respectively defined by

$$\mathcal{H}_0\{e_j\} := \sum_k \epsilon_k c_k^\dagger c_k - \Delta_\mathrm{ex} \sum_{j=1}^N e_j \cdot c_j^\dagger \hat{\boldsymbol{\sigma}} c_j, \tag{3}$$

$$\mathcal{V} := \sum_k c_k^\dagger \hat{v}_k c_k, \tag{4}$$

where $\epsilon_k := -2t(\cos k_\mathrm{x} a + \cos k_\mathrm{y} a)$ ; $\hat{v}_k := \lambda(\hat{\sigma}_\mathrm{x} \sin k_\mathrm{y} a - \hat{\sigma}_\mathrm{y} \sin k_\mathrm{x} a)$ ; $c_k$ is the annihilation operator of the electron with a wave vector $\boldsymbol{k} = (k_\mathrm{x}, k_\mathrm{y})$ in the spinor representation; $c_j := N^{-1/2} \sum_k c_k \exp(i\boldsymbol{k} \cdot \boldsymbol{R}_j)$, $\hat{\boldsymbol{\sigma}} := (\hat{\sigma}_\mathrm{x}, \hat{\sigma}_\mathrm{y}, \hat{\sigma}_\mathrm{z})$ for the Pauli matrices $\hat{\sigma}_\alpha$ (a quantity with the hat symbol is a $2 \times 2$ matrix); and $t$, $\lambda$, and $\Delta_\mathrm{ex}$ are the strengths of the hopping integral, the Rashba SOI, and the local EXF, respectively.

*Definition of* $K_\mathrm{u}^\mathrm{EXF}$. In the DLM scheme, the grand potential density is microscopically represented by [50]

$$j_{\beta\mu} = -\frac{1}{\pi V} \lim_{\eta \searrow 0} \int_{-\infty}^{\infty} dE\, \Im\, \text{Tr}\langle \mathcal{G}(E_\eta, \{e_j\})\rangle f_\beta(E-\mu)(E-\mu)$$

$$-\frac{1}{\beta\pi V} \lim_{\eta \searrow 0} \int_{-\infty}^{\infty} dE\, \Im\, \text{Tr}\langle \mathcal{G}(E_\eta, \{e_j\})\rangle \{f_\beta(E-\mu)\ln f_\beta(E-\mu) \qquad (5)$$

$$+[1-f_\beta(E-\mu)]\ln[1-f_\beta(E-\mu)]\} + \frac{1}{\beta V}\langle \ln W\{e_j\}\rangle,$$

where $E_\eta := E+i\eta$; $\mathcal{G}(z,\{e_j\}) := (z-\mathcal{H}\{e_j\})^{-1}$ for a complex energy $z$; Tr is taken over all one-electron states; $f_\beta(x) := (e^{\beta x}+1)^{-1}$ for $\beta := (k_B T)^{-1}$; $\mu$ is the chemical potential; $V := a^3 N$ is the volume of the crystal; and $\langle ... \rangle := \text{Tr}_{\{e_j\}} W\{e_j\} ... := \int de_1 de_2 \cdots de_N W\{e_j\} ...$ denotes the EXF-configuration average integrated on the unit spheres with respect to a probability distribution $W\{e_j\}$. Within the CPA, we can represent this probability distribution as [51] $W\{e_j\} = \prod_{j=1}^{N} w(e_j)$, where $w(e) := \exp[-\beta\Omega(e)]/\int de'\,\exp[-\beta\Omega(e')]$ and $\Omega(e) := \pi^{-1}\lim_{\eta \searrow 0} \Im \int_{-\infty}^{\infty} dE\, f_\beta(E-\mu)\text{Tr}_{\text{spin}} \ln\{\hat{1} - [-\Delta_{\text{ex}} e\cdot\hat{\sigma} - \hat{\Sigma}(E_\eta)]\hat{G}(E_\eta)\}$ ($\text{Tr}_{\text{spin}}$ is taken in a $2\times 2$ spin space). The CPA-Green function is given by $\hat{G}(z) = N^{-1}\sum_k [(z-\epsilon_k)\hat{1} - \hat{v}_k - \hat{\Sigma}(z)]^{-1}$, and $\hat{\Sigma}(z)$ is determined to satisfy the CPA condition [42,46]. The one-body part of the Green function in Eq. (5) is calculated using the expression $\text{Tr}\langle \mathcal{G}(z,\{e_j\})\rangle = N\text{Tr}_{\text{spin}}\hat{G}(z)$. To define the EXF-direction dependence of $j_{\beta\mu}$, we can consider a rotated EXF given by $R(\Theta,\Phi)\sum_i \langle e_i\rangle = \sum_i \text{Tr}_{\{e_j\}} W\{R(\Theta,\Phi)^{-1} e_j\} e_i$, where the EXF direction before the rotation is represented by $\sum_i \langle e_i\rangle$ and $R(\Theta,\Phi)$ is a rotation matrix with respect to the rotation angles $(\Theta,\Phi)$. Using this configuration of rotated EXFs, we define a rotated grand potential density $J_{\beta\mu}(\Theta,\Phi)$, which is expressed as

$$J_{\beta\mu}(\Theta,\Phi) := j_{\beta\mu} \text{ in replacement of } W\{e_j\} \to W\{R(\Theta,\Phi)^{-1} e_j\}, \qquad (6)$$

where the rotation matrix acting on a vector is expressed in the Cartesian coordinates as follows:

$$R(\Theta,\Phi) = \begin{pmatrix} \cos\Theta\cos\Phi & -\sin\Phi & \sin\Theta\cos\Phi \\ \cos\Theta\sin\Phi & \cos\Phi & \sin\Theta\sin\Phi \\ -\sin\Theta & 0 & \cos\Theta \end{pmatrix}. \qquad (7)$$

We introduce the angle-dependent Helmholtz energy density to fix the electron concentration to $n$. This energy density is defined by the Legendre transformation as $F_{\beta n}(\Theta,\Phi) := \max_\mu [J_{\beta\mu}(\Theta,\Phi) + \mu n]$. Further, with $\Delta F_{\beta n}(\Theta,\Phi) := F_{\beta n}(\Theta,\Phi) - F_{\beta n}(0,0)$, we define an

MAC by
$$K_\mathrm{u}^{\mathrm{EXF}}(T) := \Delta F_{\beta n}(\pi/2, 0), \tag{8}$$
where $\Delta F_{\beta n}(\pi/2, 0)$ represents the energy density needed to rotate the EXF from the easy axis to the hard axis.

*Definition of $\widetilde{K}_\mathrm{u}^{\mathrm{EXF}}$.* Perturbatively expanding $\Delta F_{\beta n}(\Theta, \Phi)$ with respect to $\lambda$, we obtain

$$\Delta F_{\beta n}(\Theta, \Phi) = -\lim_{\eta \searrow 0} \frac{1}{V} \int_{-\infty}^{\infty} \frac{dE}{2\pi} f_\beta(E - \mu_0)$$
$$\times \Im \mathrm{Tr} \left\langle [\mathcal{G}_0(E_\eta, \{e_j\}) \mathcal{U}(\Theta, \Phi)^\dagger \mathcal{V} \mathcal{U}(\Theta, \Phi)]^2 - [\mathcal{G}_0(E_\eta, \{e_j\}) \mathcal{V}]^2 \right\rangle_0 + \mathcal{O}(\lambda^4), \tag{9}$$

where quantities with the index 0 have $\lambda = 0$ in each evaluation, and $\mathcal{U}(\Theta, \Phi)$ is a unitary operator satisfying $\mathcal{U}(\Theta, \Phi) \mathcal{H}_0\{e_j\} \mathcal{U}(\Theta, \Phi)^\dagger = \mathcal{H}_0\{R(\Theta, \Phi) e_j\}$. Here, we assume that $\langle ... \rangle \simeq \langle ... \rangle_0$, neglecting the effect of $\lambda$ on the EXF structure. To calculating the two-body part of the Green functions according to Butler [52] within the CPA, we have $\mathrm{Tr} \left\langle [\mathcal{G}_0(z, \{e_j\}) \mathcal{U}(\Theta, \Phi)^\dagger \mathcal{V} \mathcal{U}(\Theta, \Phi)]^2 \right\rangle_0 = \mathrm{Tr} \left[ \langle \mathcal{G}_0(z, \{e_j\}) \rangle_0 \mathcal{U}(\Theta, \Phi)^\dagger \mathcal{V} \mathcal{U}(\Theta, \Phi) \right]^2$. This indicates that the vertex correction vanishes because of the symmetry of the Rashba SOI, i.e., $\hat{v}_{-k} = -\hat{v}_k$. Subsequently, the explicit form of the angle dependence is obtained as

$$\mathrm{Tr} \left\langle [\mathcal{G}_0(z, \{e_j\}) \mathcal{U}(\Theta, \Phi)^\dagger \mathcal{V} \mathcal{U}(\Theta, \Phi)]^2 - [\mathcal{G}_0(z, \{e_j\}) \mathcal{V}]^2 \right\rangle_0$$
$$= \sin^2 \Theta \, \lambda^2 [\Sigma_0^+(z) - \Sigma_0^-(z)]^2 \sum_k [g_k^+(z) g_k^-(z)]^2 \sin^2 k_x a, \tag{10}$$

where $\Sigma_0^\sigma(z)$ and $g_k^\sigma(z)$ are the $\sigma$-spin elements in $\hat{\Sigma}_0(z)$ and $\hat{g}_k(z) := [(z - \epsilon_k)\hat{1} - \hat{\Sigma}_0(z)]^{-1}$, respectively (these are diagonal). Thus, from the usual definition of the first-order MAC of $\Delta F_{\beta n}(\Theta, \Phi) = \widetilde{K}_\mathrm{u}^{\mathrm{EXF}} \sin^2 \Theta + \mathcal{O}(\lambda^4)$, we have

$$\widetilde{K}_\mathrm{u}^{\mathrm{EXF}}(T) = -\lim_{\eta \searrow 0} \Im \int_{-\infty}^{\infty} \frac{dE}{2\pi} f_\beta(E - \mu_0)$$
$$\times [\Sigma_0^+(E_\eta) - \Sigma_0^-(E_\eta)]^2 \frac{\lambda^2}{V} \sum_k [g_k^+(E_\eta) g_k^-(E_\eta)]^2 \sin^2 k_x a. \tag{11}$$

We now present a power law for the first-order MAC within the perturbative treatment. We introduce a (nonperturbative) magnetization as

$$\widetilde{M}(T) := -\frac{g\mu_B}{2} \lim_{\eta \searrow 0} \Im \int_{-\infty}^{\infty} \frac{dE}{\pi} f_\beta(E - \mu_0) \tag{12}$$

$$\times [\Sigma_0^+(E_\eta) - \Sigma_0^-(E_\eta)] \frac{1}{V} \sum_{\boldsymbol{k}} g_{\boldsymbol{k}}^+(E_\eta) g_{\boldsymbol{k}}^-(E_\eta),$$

where $g$ and $\mu_B$ are the g factor of an electron and the Bohr magneton, respectively. When comparing Eqs. (11) and (12), using the Born approximation $\Sigma_0^\sigma(z) \simeq -\sigma \Delta_{\mathrm{ex}} \langle e_z \rangle + G_0^\sigma(z) \Delta_{\mathrm{ex}}^2 (1 - \langle e_z \rangle^2)$ and then approximating as $\Sigma_0^+(E_\eta) - \Sigma_0^-(E_\eta) \simeq -2\Delta_{\mathrm{ex}} \langle e_z \rangle + \mathcal{O}(\Delta_{\mathrm{ex}}^2)$, we obtain $\widetilde{K}_u^{\mathrm{EXF}}(T)/\widetilde{K}_u^{\mathrm{EXF}}(0) \simeq \langle e_z \rangle^2$ and $\widetilde{M}(T)/\widetilde{M}(0) \simeq \langle e_z \rangle$. Thus, a second power law is obtained as

$$\frac{\widetilde{K}_u^{\mathrm{EXF}}(T)}{\widetilde{K}_u^{\mathrm{EXF}}(0)} \simeq \left(\frac{\widetilde{M}(T)}{\widetilde{M}(0)}\right)^2, \tag{13}$$

within the perturbative treatment up to the order of $\lambda^2 \Delta_{\mathrm{ex}}$. Outside the perturbative range, this second power law is not necessarily valid, as explained below.

First, we numerically exam the power law between $\widetilde{K}_u^{\mathrm{EXF}}(T)$ and $\widetilde{M}(T)$. In the perturbative theory with respect to $\lambda$, the model parameters are $n$ and $\Delta_{\mathrm{ex}}/t$. We set electron concentration to $n = 0.27$, which satisfies the condition of ferromagnetism with the uniaxial MA. Figure 1 shows the temperature dependences of $\widetilde{k}_u^{\mathrm{EXF}}(T) := \widetilde{K}_u^{\mathrm{EXF}}(T)/\widetilde{K}_u^{\mathrm{EXF}}(0)$ and $\tilde{\alpha}(T) := \ln \widetilde{k}_u^{\mathrm{EXF}}(T) / \ln \widetilde{m}(T)$, where $\widetilde{m}(T) := \widetilde{M}(T)/\widetilde{M}(0)$, and $\widetilde{m}(T)^2$, $\widetilde{m}(T)^3$, and $\widetilde{m}(T)^4$ are shown as references for observing the power law. For $\Delta_{\mathrm{ex}}/t = 0.2$ [Fig. 1(a)], the second power law [$\tilde{\alpha}(T) \simeq 2$] can be clearly observed, but $\tilde{\alpha}(T)$ and its temperature dependency are found to increase with increasing $\Delta_{\mathrm{ex}}/t$ [Fig. 1(b)-(d)]. From these results, we suggest that the ratio of the exchange-splitting energy divided by the band width is an important factor for establishing the second power law in metallic ferromagnetism. In other words, we can consider that this tendency is consistent with the physically observed fact that itinerant systems are likely to exhibit $\tilde{\alpha} = 2$ whereas localized ones exhibit $\tilde{\alpha} > 2$. As shown in Fig. 1(d), the exponent $\tilde{\alpha}$ exceeds the value 3 in the localized electron limit. At this point, one can compare the spin model and the present model. When considering a spin Hamiltonian density including the MA term of $-\sum_i K_i (S_i^z)^2 - \sum_{i \neq j} K_{ij} S_i^z S_j^z$, the Callen–Callen theory [12] predicts the exponent $\alpha_{\mathrm{CC}} = (2K_{\mathrm{int}} + 3K_{\mathrm{on}})/(K_{\mathrm{int}} + K_{\mathrm{on}})$ in the low-temperature limit [53]. Here, $K_i$ and $K_{ij}$ respectively represent single-ion and two-ion anisotropies; $\boldsymbol{S}_i$ is the spin operator with a spin $S$ at site $i$; $K_{\mathrm{on}} := S(S - 1/2) \sum_i K_i$; and $K_{\mathrm{int}} := (S^2/2) \sum_{i \neq j} K_{ij}$. From this expression for $\alpha_{\mathrm{CC}}$, we obtain $2 < \alpha_{\mathrm{CC}} \leq 3$ for $K_{\mathrm{int}}/K_{\mathrm{on}} > 0$, as expected. Furthermore, $3 < \alpha_{\mathrm{CC}}$ for $K_{\mathrm{int}}/K_{\mathrm{on}} < 0$ [54]. These findings for the spin system indicate that itinerant ferromagnets, which involve both spin and orbital degrees of freedom, have the potential to realize a wide range of $\tilde{\alpha}$ other than $2 \leq \tilde{\alpha} \leq 3$, even with changing $\Delta_{\mathrm{ex}}/t$.

Next, we consider an equality of the formulations of $K_u^{\mathrm{MP}}(T)$ and $K_u^{\mathrm{EXF}}(T)$. Although both of them estimate the MA energy density of a material, their theoretical concepts are different. $K_u^{MP}(T)$ is defined as the energy density required to magnetize by an external magnetic field along the hard axis. Accordingly, it is expressed by the magnetic susceptibility as shown in Eq. (1). In contrast, $K_u^{EXF}(T)$ is defined as the energy density required to virtually rotate the EXF from the easy axis to the hard axis in a fixed spin structure. Figure 2 shows four cases: (a) weak EXF and weak SOI, (b) strong EXF and weak SOI, (c) weak EXF and strong SOI, and (d) strong EXF and strong SOI. The values of EXF and SOI are provided in the caption of Fig. 2. In addition to the MACs, we also plot the exponent calculated from $K_u^{EXF}(T)$, which is defined by $\alpha(T) \coloneqq \ln[K_u^{EXF}(T)/K_u^{EXF}(0)]/\ln[M(T)/M(0)]$, where $M(T) \coloneqq -\frac{g\mu_B}{2a^3}\lim_{\eta \searrow 0} \Im \int_{-\infty}^{\infty} \frac{dE}{\pi} f_\beta(E-\mu) \operatorname{Tr}_{\text{spin}} \hat{\sigma}_z \hat{G}(E_\eta)$. As expected, all MACs match well in the weak-$\lambda$ cases (a) and (b), and there is a large difference between $K_u^{EXF}(T)$ and $\widetilde{K}_u^{EXF}(T)$ in the strong-$\lambda$ cases (c) and (d). Although $K_u^{MP}(T)$ and $K_u^{EXF}(T)$ seem to yield almost the same result in the all cases, $K_u^{MP}(T)$ slightly deviates from $K_u^{EXF}(T)$ in the strong-$\lambda$ cases. This deviation implies that $K_u^{MP}(T)$ and $K_u^{EXF}(T)$ are not always consistent. In particular, although not shown in this letter, we observe that the difference between them at the zero temperature increases when the chemical potential exists at a highly half-metallic region in the density of states in the electronic structure. These temperature dependences differ at higher temperatures, indicating that the half metallicity is lost as the temperature increases. In addition, a noncollinear effect (NCE) between the directions of the magnetization and the EXF yields a mismatch between $K_u^{MP}(T)$ and $K_u^{EXF}(T)$ in general, as shown in our previous work on the MA of Dy-Fe-B magnets [7] (in the present conditions, we have confirmed that the NCE is negligible). Because such special behaviors may strongly depend on theoretical models, these cases should be studied in future work.

In summary, we theoretically investigated the temperature-dependent MA in ferromagnetic metals with a uniaxial MA using the Rashba model. We examined the temperature dependences of $K_u^{MP}$, $K_u^{EXF}$, and $\widetilde{K}_u^{EXF}$ based on the Sucksmith–Thompson theory, the EXF-rotation concept, and the perturbative theory with respect to the Rashba SOI, respectively. In addition, we demonstrated the validity of the power law. Consequently, we confirmed that $\widetilde{K}_u^{EXF}$ functions well for a weak SOI and helps determine the effect of the strength of the EXF on its temperature dependence. This enhances the understanding of the second power law observed in ferromagnetic metals with a uniaxial MA. $K_u^{MP}(T)$ and $K_u^{EXF}(T)$ can also be used in a strong-SOI case to estimate the MA, and the values match well in the present conditions.

However, they are not always consistent, and further investigations are required to study this discrepancy.

**Acknowledgment**

This work was supported by JSPS KAKENHI Grant Numbers JP17K14800, JP19H05612, and JP21K04624 in Japan.

*E-mail: dmiura1222@gmail.com

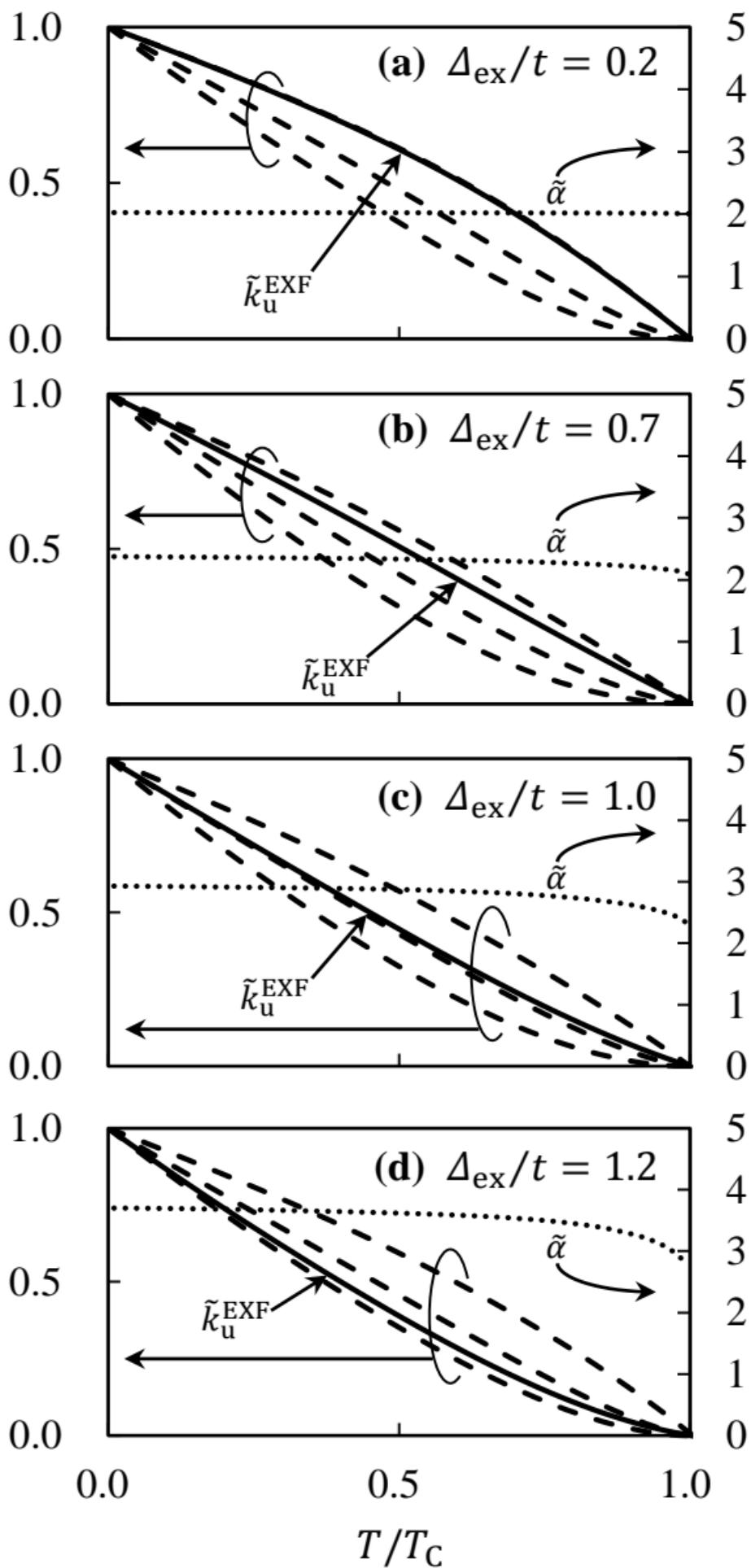

**Fig. 1** Calculated temperature dependences of $\tilde{k}_{\mathrm{u}}^{\mathrm{EXF}}(T)$ and $\tilde{\alpha}(T)$, denoted by the solid and dotted lines, respectively. The upper, middle, and lower dashed lines in each figure represent $\tilde{m}(T)^2$, $\tilde{m}(T)^3$, and $\tilde{m}(T)^4$, respectively.

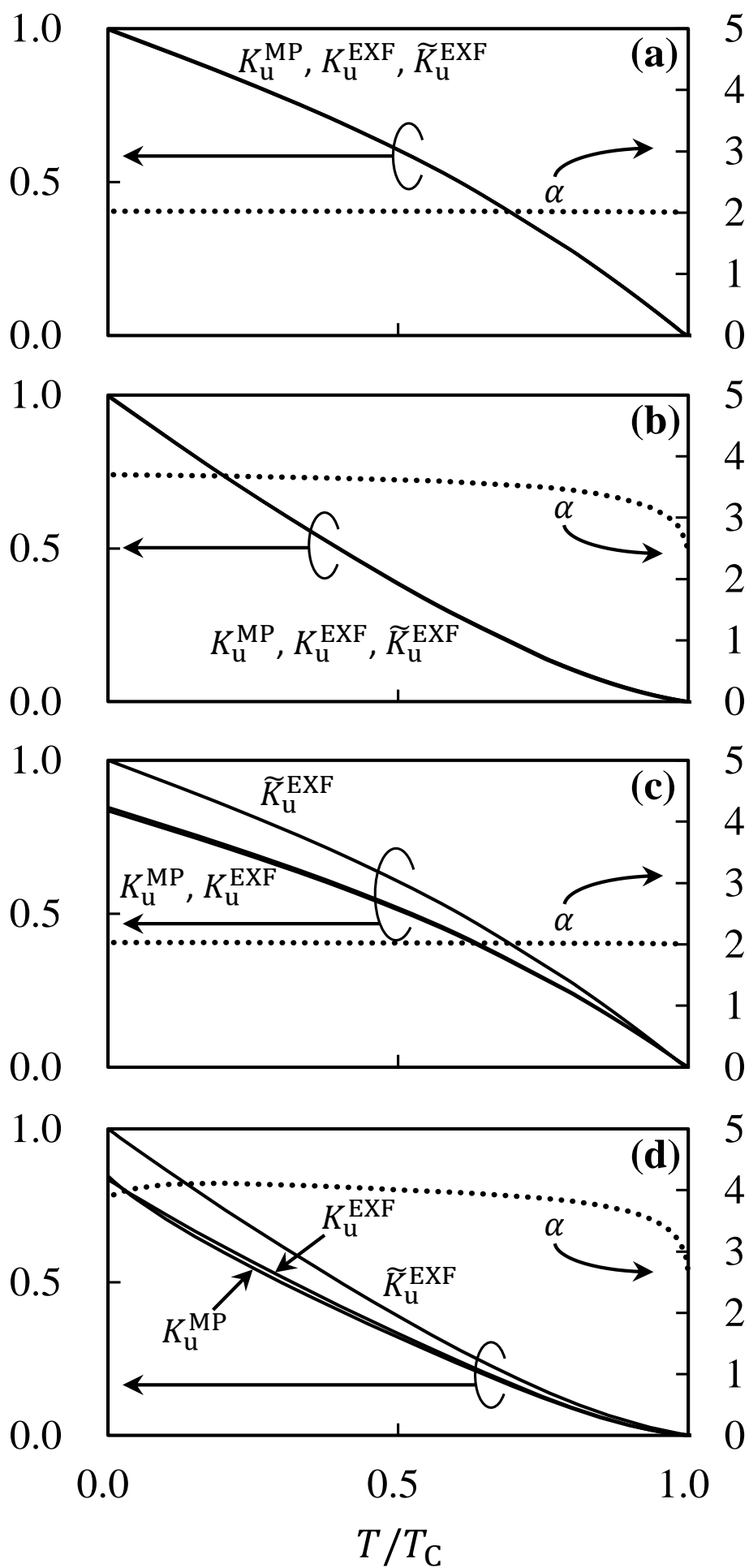

**Fig. 2** Calculated temperature dependences of $K_u^{MP}(T)$, $K_u^{EXF}(T)$, $\widetilde{K}_u^{EXF}(T)$, and $\alpha(T)$. The solid and dotted lines represent the MACs and $\alpha(T)$, respectively. In each case, the MACs are normalized by $\widetilde{K}_u^{EXF}(0)$. The model parameters, $(\Delta_{ex}, \lambda)/t$, are set as (a) (0.2, 0.1), (b) (1.2, 0.1), (c) (0.2, 0.8), and (d) (1.2, 0.8).